\documentclass[onecolumn,english,showpacs,preprintnumbers,amsmath,amssymb,floatfix]{revtex4}
\usepackage[T1]{fontenc}
\usepackage[latin9]{inputenc}
\usepackage{color}
\usepackage{array}
\usepackage{amstext}
\usepackage{graphicx}
\usepackage{esint}

\makeatletter

\@ifundefined{textcolor}{}
{%
 \definecolor{BLACK}{gray}{0}
 \definecolor{WHITE}{gray}{1}
 \definecolor{RED}{rgb}{1,0,0}
 \definecolor{GREEN}{rgb}{0,1,0}
 \definecolor{BLUE}{rgb}{0,0,1}
 \definecolor{CYAN}{cmyk}{1,0,0,0}
 \definecolor{MAGENTA}{cmyk}{0,1,0,0}
 \definecolor{YELLOW}{cmyk}{0,0,1,0}
 }

\@ifundefined{definecolor}
 {\usepackage{color}}{}
\@ifundefined{definecolor}
 {\usepackage{color}}{}
\makeatother

\makeatother

\usepackage{babel}

\begin{document}
\title{Heavy quark fragmentation functions at next-to-leading perturbative QCD}
\author{S. M. Moosavi Nejad$^{a,b}$}
\author{P. Sartipi Yarahmadi$^{a}$}
\email{mmoosavi@yazd.ac.ir}

\affiliation{$^{(a)}$Faculty of Physics, Yazd University, P.O. Box
89195-741, Yazd, Iran}

\affiliation{$^{(b)}$School of Particles and Accelerators,
Institute for Research in Fundamental Sciences (IPM), P.O.Box
19395-5531, Tehran, Iran}

\date{\today}

\begin{abstract}

It is well-known that the dominant mechanism to produce hadronic bound states
with large transverse momentum is fragmentation. 
This  mechanism is described by the fragmentation functions (FFs) which are the universal and process-independent functions.
Here, we review the perturbative FFs formalism as an appropriate
tool for studying these hadronization processes and detail the extension of this formalism at next-to-leading order (NLO).	
Using the Suzuki's model, we calculate the perturbative QCD FF for a heavy quark to fragment into a S-wave heavy meson at NLO.
As an example, we study  the  LO and NLO FFs for a charm quark to split into the S-wave $D$-meson and compare our analytic results both
with experimental data  and well-known phenomenological models. 
\end{abstract}


\maketitle

\section{Introduction}
\label{sec:intro}

Heavy quark production processes provide a powerful insight into our 
understanding of Quantum Chromodynamics (QCD). 
The study of heavy mesons properties is also a subject of 
interest for understanding of quark-gluon interaction dynamics.
Generally, two mechanisms are investigated for the production of heavy mesons:  recombination and fragmentation \cite{Martynenko:1995hg}. In the first scheme, heavy mesons are formed from heavy-heavy or heavy-light quarks which are produced independently in hard subprocesses. In the second mechanism, the fragmentation refers to the process of a parton which carries large transverse momentum and subsequently forms a jet containing the expected hadron \cite{Braaten:1993rw}. 
At sufficiently large transverse momentum of the heavy meson production, the direct leading-order production scheme (recombination mechanism)
is normally suppressed while the fragmentation scheme becomes dominant, though it is formally of higher order
in the strong coupling constant $\alpha_s$ \cite{Braaten:1993rw,Kramer:2001hh}.\\
The fragmentation mechanism is described by the 
 function $D_i^M(z,\mu_0)$ which refers to the probability for a 
parton $\it{i}$ at the factorization scale $\mu_0$ to fragment into a hadron $\it{M}$ carrying away a fraction $\it{z}$ of its momentum \cite{Braaten:1993rw}. The fragmentation functions (FFs)  are key quantities to calculate hadroproduction cross sections
and their specific importance   is for their model-independent predictions of the cross sections at the Large Hadron Collider (LHC). In this respect, one needs to determine these functions with high accuracy as possible.\\
Basically, there are two approaches to determine the FFs where one calculates these functions in the initial scale of fragmentation and then can evolve them to higher scales using the
Dokshitzer-Gribov-Lipatov-Altarelli-Parisi (DGLAP) 
renormalization group equations \cite{dglap}. 
In  the first approach (\textit{phenomenological} approach) the free parameters in the proposed forms of the FFs are extracted form experimental data analysis. Since the hadronization mechanism is universal and independent of the perturbative processes which produce the partons  one can exploit, for example, the existing data on $e^+e^-\rightarrow i\bar{i}\rightarrow M+jets$ events to fit the proposed models for the $i\rightarrow M$ transition. In \cite{Soleymaninia,Soleymaninia:2014oya}, considering a power model for the FFs we determined the $\pi^\pm/K^\pm$ FFs, both at LO and NLO,  through a global fit to the single-inclusive $e^+e^-$ annihilation data and the semi-inclusive deep inelastic scattering asymmetry data from HERMES and COMPASS. This phenomenological approach is frequently used to obtain the nonperturbative FFs.
  
The FFs are related to the low-energy part of the hadroproduction processes but, fortunately, it was
found that these functions for heavy hadron productions are analytically calculable by virtue of perturbative QCD (pQCD) with limited phenomenological 
parameters \cite{Ma:1997yq,Chang:1991bp,Braaten:1993mp}.
An alternative fragmentation model which does contain spin information has been proposed by Suzuki \cite{Suzuki:1977km,Suzuki:1985up}. 
In this approach, Suzuki calculates the heavy FFs using the same  Feynman diagrams, as in the pQCD approach, for the parton level of the process and also by considering  the wave function of heavy meson which
 contains the bound state nonperturbative dynamic of
produced meson. In the Suzuki's model, the 
analytical expression of FFs depends on the transverse momentum $k_T$ of the initial parton, while in the pQCD scheme one integrates  over the invariant
mass of the fragmenting quark. The invariant mass is related to the transverse
momentum $k_T$ of the meson relative to the fragmenting quark \cite{Braaten:1994bz}. In fact, rather than integrating over $k_T^2$, the Suzuki's model chooses  to evaluate the integrand at a typical value $\left\langle k_T^{2}\right\rangle$, see (\ref{ayda2}).
In \cite{Nejad:2013vsa,MoosaviNejad:2016scq}, using the Suzuki's model
we calculated the initial scale fragmentation function for  $\it{c}$-quark to split into
S-wave $D^0/D^+$-mesons at leading-order of perturbative QCD.
Here, we review the Suzuki's formalism at LO 
and detail the extension of this formalism at NLO by considering the real and virtual gluon radiative corrections. 
Finally we will present, for the first time, our NLO analytical expression of the heavy quark FF  in the Suzuki's model and compare
our result with the LO one \cite{Nejad:2013vsa}.
To show the importance of our calculations we will also compare both results with experimental data from BELLE \cite{belle} and CLEO \cite{cleo}. We also compare our analytical results with a well-known phenomenological model.
As will be shown, the NLO corrections improve our theoretical results at LO and
make good agreement with  experimental data.

\section{Calculation of fragmentation function at  NLO perturbative QCD}
\label{sec:one}

\begin{figure}
	\begin{center}
		\includegraphics[width=0.35\linewidth,bb=180 470 452 709]{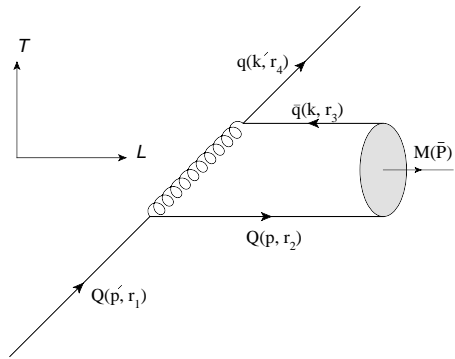}
		\caption{\label{lo}%
			The lowest order Feynman diagram contributing to the fragmentation of a heavy quark Q
			into a heavy meson $M(Q\bar{q})$.}
	\end{center}
\end{figure}

The theoretical approaches to calculate the heavy quark FFs depend on the fact that 
the FFs for hadrons containing heavy quarks can be calculated theoretically using the
perturbative QCD (pQCD) \cite{Braaten:1993rw}.
The first theoretical effort to illustrate the production procedure of hadrons containing  heavy quarks
was established by Bjorken \cite{Bjorken:1977md}. In a
naive quark-parton model, he deduced that the inclusive distribution of  heavy mesons  should peak
almost at $z=1$, where $z$ refers to the longitudinal momentum fraction of the hadron state. 
The pQCD scheme  was followed  by  Suzuki \cite{Suzuki:1977km},  Ji and Amiri \cite{Amiri:1986zv} by
considering more elaborate models. While
in this approach Suzuki computes  the heavy FFs by applying a Feynman diagram similar to that in Fig.~\ref{lo},
Amiri and Ji calculate their FFs in  $e^+e^-$ annihilation process in the same order of pQCD.
In their models the total amplitude for the fragmentation of a heavy quark into a heavy meson is obtained
by perturbative calculations of quark-gluon non-Abelian interaction up to the order
of $\alpha_s^2$ and use of a delta function to represent the S-wave heavy meson bound state.
In fact, they consider a heavy bound state as a nonrelativistic system and reduce its
wave function to a delta function \cite{Brodsky:1985cr}.

\begin{figure}
	\begin{center}
		\includegraphics[width=0.5\linewidth,bb=188 430 452 709]{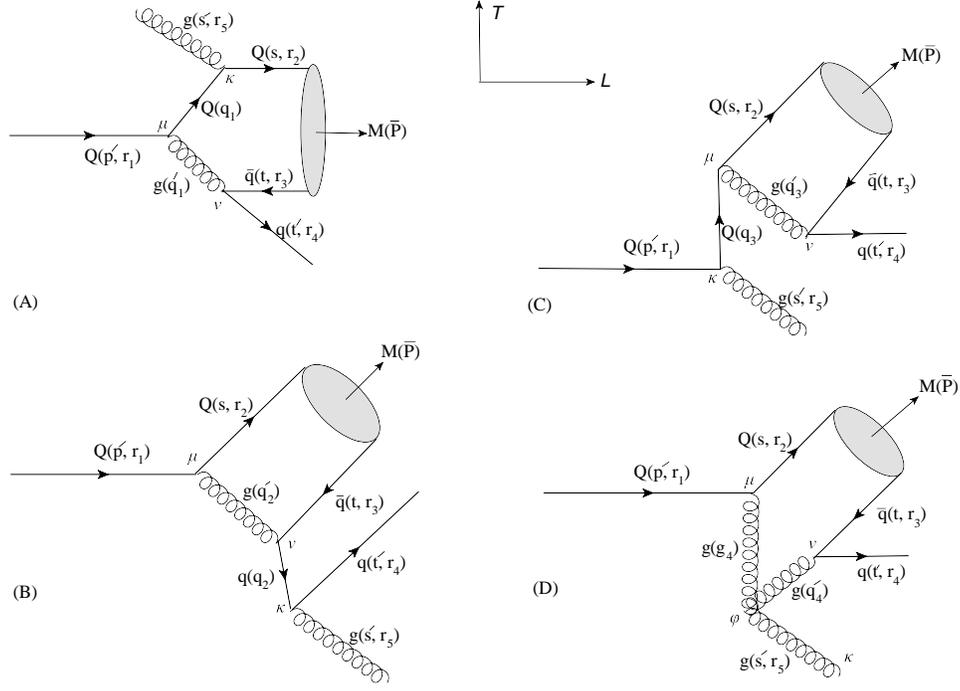}
		\caption{\label{nlo}%
			Production of a heavy meson $M(Q\bar{q})$ at next-to-leading order (NLO). 
			Real gluon radiative contributions to $Q\rightarrow M(Q\bar{q})+q$ are shown at NLO.
			The spins ($r_i$) and  four-momenta are also labeled.}
	\end{center}
\end{figure}
 The Suzuki's model includes most of the kinematical and dynamical properties of the splitting process and gives us a detailed insight  about the fragmentation process.
 Specifically, if  we want to know about further details
 such as the spin property of fragmentation, this model can be instructive. Briefly, the Suzuki's model is a dynamical model which is  more predictive and describes spin-dependent effects and also includes the kinematical details of the fragmentation process. 
 It mixes a perturbative picture with nonperturbative dynamics of fragmentation and not only predicts the z-dependence of the FFs, but also their dependence on $k_T$, the transverse momentum of the meson relative to the jet.\\
In \cite{Nejad:2013vsa}, using the Suzuki's model we derived an analytical expression 
for the heavy quark FF at lowest order ($\alpha_s^2$-order) of pQCD by considering the typical Feynman 
diagram  shown in Fig.~\ref{lo},
where a heavy quark Q creates a bound state $M(Q\bar{q})$ along with a light quark $q$ through a single gluon.
The result for the fragmentation function $D_{Q\rightarrow M}(z, \mu_0)$ 
 was dependent on the transverse momentum $k_T$ of the initial heavy quark.
Here, we present  a compacted expression of our previous result as
\begin{eqnarray}\label{last}
D_{Q\rightarrow M}^{LO}(z, \mu_0)&=&\frac{2B^2 \alpha_s^2}{3}C_F^2\frac{z(z-1)^3}{F(z, \left\langle k_T^{2}\right\rangle)}
\bigg\{\frac{m_Q M^4}{m_q}(1-z)^2  +2\frac{z(1-z)M^3}{m_q}\bigg(m_q m_Q-(1-z)m_Q^2 \bigg)
\nonumber\\
&&+\frac{m_QM^2}{m_q}z^2\bigg((2+3z^2-2z)\left\langle k_T^{2}\right\rangle+3m_q^2
+3m_Q^2(1-z)^2-8m_qm_Q(1-z)\bigg)
\nonumber\\
&&-2\frac{m_QM}{m_q}z^3\bigg(\big[m_Q-(1-z)m_q\big]\left\langle k_T^{2}\right\rangle+m_qm_Q\big[m_q-(1-z)m_Q\big]\bigg)
\nonumber\\
&&+\big[\left\langle k_T^{2}\right\rangle+m_q^2\big]\big[\left\langle k_T^{2}\right\rangle+m_Q^2\big]\frac{m_Qz^4}{m_q}\bigg\},
\end{eqnarray}
where,
\begin{eqnarray}
F(z, \left\langle k_T^{2}\right\rangle)&=&\bigg[(z-1)(M^2-zm_Q^2)-z(m_q^2+z\left\langle k_T^{2}\right\rangle)\bigg]^2\times
\bigg[z^2 \left\langle k_T^{2}\right\rangle+(M(z-1)-zm_q)^2\bigg]^2,
\end{eqnarray}
and $B=\pi m_Q m_{\bar q}f_M$, but the coefficient B is related to the normalization condition
$\int_0^1 D_{Q\rightarrow M}(z, \mu_0) dz=1$  \cite{Amiri:1986zv,Suzuki:1985up}.
In computing Eq.~(\ref{last}), following Ref.~\cite{Suzuki:1985up} we adopted the infinite momentum frame where
the fragmentation  parameter in the usual light-cone form, $z=(p^0_M+p^3_M)/(p^0_Q+p^3_Q)$, is 
reduced to a more popular form $z=p^0_M/p^0_Q=E_M/E_Q$ which is more convenient when the masses of partons and outgoing meson are ignored. 
In reality, the scaling variable $z$ refers to 
the energy fraction of the fragmenting heavy quark which is taken away by the produced meson and takes the values as $0\le z \le 1$.
In \cite{Qi:2007sf}, authors studied
the theoretical uncertainties due to the freedom in the choice of fragmentation parameter in the presence of heavy quark and meson masses.
In fact, hadron mass modifies the relations between
partonic and hadronic variables and  is responsible for the low-$z$ threshold, although this additional effect is
not expected to be truly sizable numerically, its study is nevertheless necessary in order to fully exploit the enormous
statistics of the LHC data.

In the present work we derive an analytical form
of the transverse momentum dependent heavy quark FF 
at next-to-leading order (NLO) with assumption of a delta function for the meson bound state, as in \cite{Amiri:1985mm}.
The underlying link between hadronic phenomena in QCD at large and small distances is the hadronic wave function. 
In fact, the nonperturbative aspect of the hadroprodution processes is contained in the bound state of the meson
which is described by the wave function.
Following Ref.~\cite{Suzuki:1985up} and according to the Lepage-Brodsky's approach \cite{Lepage:1980fj} we 
neglect the relative motion of  the constituent quarks $Q$ and $\bar q$
therefore we assume, for simplicity, that the quark pair $Q\bar{q}$
are emitted collinearly with each other and  move  along the $Z$-axes.
In fact, in the Suzuki's model a meson is replaced by collinear
constituents with neglecting the Fermi motion and the nonperturbative aspect of the hadroproduction
is included  in the wave function of the heavy meson bound state.\\
The Feynman diagrams of the real gluon corrections are shown in Fig.~\ref{nlo}.
Considering these diagrams we set the relevant four-momenta  as
\begin{eqnarray}\label{kinematic}
p_\mu^\prime =[p_0^\prime, \vec{k}_T, p_L^\prime], &&\quad      s_\mu=[s_0, \vec{0}, s_L], \nonumber\\
s_\mu^\prime =[s_0^\prime, \vec{s^\prime}_T, s_L^\prime],  &&\quad     t_\mu=[t_0, \vec{0}, t_L],\\
t_\mu^\prime =[t_0^\prime, \vec{t^\prime}_T, t_L^\prime],  &&\quad     \bar P_\mu=[\bar P_0, \vec{0}, \bar P_L],\nonumber
\end{eqnarray}
where $\bar P$ refers to the four-momentum of the produced meson, so  $ \bar{P}_L=s_L+t_L$.
\\
To proceed, we define the momentum fractions carried by the constituent quarks as: $x_1=(s_0+s_L)/(\bar{P}_0+\bar{P}_L)$
and $x_2=(t_0+t_L)/(\bar{P}_0+\bar{P}_L)$ so that $x_1+x_2=1$. In the infinite momentum frame these fractions are 
reduced to simpler forms as
\begin{eqnarray}\label{zare}
x_1=\frac{s_0}{\bar{P}_0}\quad,\quad x_2=\frac{t_0}{\bar{P}_0}\cdot
\end{eqnarray}
Thus, $x_1$ and $x_2$ stand for the meson energy fractions carried by the constituent quarks.
Considering the definition of fragmentation  parameter, $z=E_M/E_Q=\bar P_0/p_0^\prime$, we
also may write the parton energies in terms of the  initial heavy quark energy $p_0^\prime$  as
\begin{eqnarray}
s_0=x_1 z p_0^\prime,\quad t_0=x_2 z p_0^\prime,\quad  s_0^\prime\simeq t_0^\prime=\frac{1-z}{2} p_0^\prime.
\end{eqnarray}
In the Suzuki's model the fragmentation function for the production of a S-wave 
heavy meson $M$ in the fragmentation of a quark Q may be put in the following form \cite{Suzuki:1977km,Suzuki:1985up}
\begin{eqnarray}\label{first}
D_{Q\rightarrow M}(z, \mu_0)=
\frac{1}{1+2r_1}\sum_{r_i, c_i}\int d^3\vec{\bar P} d^3\vec{t^\prime}d^3\vec{s^\prime} |T_M|^2 \delta^3(\vec{\bar P}+\vec{t^\prime}
+\vec{s^\prime}-\vec{p^\prime}),
\end{eqnarray}
where, $\mu_0$ is the fragmentation scale, $r_1$ refers to the spin of the fragmenting quark and 
the summation is going over the spins and colors.
In (\ref{first}), $T_M$ is
 the  probability amplitude of the meson production  which, at the large momentum transfer,
is expressed in terms of the  hard scattering 
amplitude $T_H$ and the process-independent distribution amplitude $\Phi_M$ \cite{brodsky,Amiri:1985mm} as 
\begin{eqnarray}\label{base}
T_M(\bar P, s^\prime, t^\prime)=\int [dx_i] T_H(\bar P, s^\prime, t^\prime, x_i) \Phi_M(x_i, Q^2),
\end{eqnarray}
where $[dx_i]=dx_1dx_2\delta(1-x_1-x_2)$. This scheme is convenient to absorb the soft behavior of the
bound state into the hard scattering amplitude \cite{Brodsky:1985cr}.
The short-distance coefficient $T_H$ can be calculated perturbatively from quark-gluon subprocesses at LO or NLO approximations.
The long-distance distribution amplitude $\Phi_M$ which contains the bound state nonperturbative dynamic of the
outgoing meson, is the probability amplitude for a $Q\bar q$-pair  to evolve
into a particular bound state.
The distribution amplitude $\Phi_M$ is related to the valence wave function of meson $\Psi_M$ \cite{brodsky}.
With the heavy quark mass, the relative motion of the constituent quarks
inside the heavy meson is effectively nonrelativistic and this allows 
one to estimate the nonrelativistic mesonic wave function as
a delta function form.
Therefore, the distribution amplitude for a S-wave heavy meson with neglecting the Fermi 
motion reads \cite{GomshiNobary:1994eq}
\begin{eqnarray}\label{delta}
\Phi_M\approx\frac{f_M}{2\sqrt{3}} \delta(x_1-\frac{m_Q}{M}),
\end{eqnarray}
where $M=m_Q+m_{\bar q}$ stands for the meson mass in the nonrelativistic limit,  $f_M=(6b^3/\pi M)^{1/2}$ refers to the decay constant of meson which
can be also related to the nonrelativistic mesonic S-wave function $\psi(0)$ at the origin as $f_M=\sqrt{12/M}|\psi(0)|$.
In the meson decay constant, $b$ is the binding energy of the mesonic bound state.
In \cite{MoosaviNejad:2016scq}, we studied the effect of  meson wave function on the heavy quark FF by considering
a typical mesonic wave function which is different of the delta function 
and is the nonrelativistic limit of the solution of Bethe-Salpeter equation  with the QCD kernel \cite{brodsky}.
However, due to the lengthy and cumbersome expressions of the new FF
we just presented the two-dimensional integrals which must be evaluated numerically.\\
Here, for simplicity, we consider a delta function  for the hadron bound state as well. With this  approximation (\ref{delta}), we  are assuming that 
the contribution of each constituent quark from the meson energy is proportional
to its mass, i.e. $x_1=m_Q/M$ and $x_2=m_{\bar q}/M$ (\ref{zare}) so that $x_1+x_2=1$.

Using Eqs.~(\ref{base}) and (\ref{delta}), one has
\begin{eqnarray}\label{pepe}
T_M(\bar P, s^\prime, t^\prime)=\frac{f_M}{2\sqrt{3}} T_H(\bar P, s^\prime, t^\prime, x_1=\frac{m_Q}{M}, x_2=\frac{m_{\bar q}}{M}).
\end{eqnarray}
Considering the NLO Feynman diagrams shown in Fig.~\ref{nlo}, where a produced meson is replaced by collinear
constituent quarks, we make the NLO approximation for the fragmentation function of $M(Q\bar q)$-meson.
In (\ref{pepe}), the QCD amplitude $T_H$ is, in essence, the partonic cross section to produce a quark
pair $Q\bar{q}$ with certain quantum number that in the old fashioned perturbation theory is expressed as
\begin{eqnarray}\label{forth}
T_H=\frac{g_s^3m_Q m_{\bar q}}{2\sqrt{2\bar{P}_0  s_0^\prime t_0^\prime p_0^\prime}}C_F\frac{\sum_{i=1}^4{\Gamma_i}}{\bar{P}_0
+s_0^\prime+t_0^\prime-p_0^\prime},
\end{eqnarray}
where, $C_F=2\sqrt{2}/3$ is the color factor for the process $Q\rightarrow M(Q\bar{q})+q+g$. In (\ref{forth}), the amplitudes $\Gamma_i$ stand for each Feynman diagrams shown in Fig.~\ref{nlo}, and include  an appropriate 
combination of the quark propagators and the spinorial parts of the amplitude. We set the amplitudes $\Gamma_1$ for Fig.~\ref{nlo}A, $\Gamma_2$ for Fig.~\ref{nlo}B, $\Gamma_3$ for Fig.~\ref{nlo}C and the amplitude $\Gamma_4$ for Fig.~\ref{nlo}D. These amplitudes read
\begin{eqnarray}\label{fifth}
\Gamma_1&=&\frac{-\epsilon_\kappa^\star}{G_1(q_1^{\prime 2}) G_2(q_1^2)}
\Big\{\bar{u}(s,r_2)\gamma^\kappa (\displaystyle{\not}q_1+m_Q)\gamma^\mu u(p^\prime,r_1)\Big\}\Big\{\bar{u}(t^\prime,r_4)\gamma_\mu v(t,r_3)\Big\},\nonumber\\
\Gamma_2&=&\frac{-\epsilon_\kappa^\star}{G_3(q_2^{\prime 2}) G_4(q_2^2)}
\Big\{\bar{u}(s,r_2)\gamma^\mu u(p^\prime,r_1)\Big\}\Big\{\bar{u}(t^\prime,r_4)\gamma^\kappa 
(\displaystyle{\not}{q}_2+m_q)\gamma_\mu v(t,r_3)\Big\},\\
\Gamma_3&=&\frac{-\epsilon_\kappa^\star}{G_5(q_3^{\prime 2}) G_6(q_3^2)}
\Big\{\bar{u}(s,r_2)\gamma^\mu(\displaystyle{\not}q_3+m_Q)\gamma^\kappa u(p^\prime,r_1)\Big\}\Big\{\bar{u}(t^\prime,r_4)\gamma_\mu v(t,r_3)\Big\},\nonumber\\
\Gamma_4&=&\frac{-i\epsilon_\kappa^\star}{G_7(q_4^{\prime 2}) G_8(q_4^2)}
\Big\{\bar{u}(s,r_2)\gamma^\mu g_{\mu\nu}\big[g_{\phi\nu}(-2s^\prime-q^\prime)_\kappa+g_{\nu\kappa}(2q^\prime+s^\prime)_\phi+g_{\kappa\phi}(s^\prime-q^\prime)_\nu\big]u(p^\prime,r_1)\Big\}\times\nonumber\\
&&\hspace{3cm}\Big\{\bar{u}(t^\prime,r_4)\gamma_\nu v(t,r_3)\Big\},\nonumber
\end{eqnarray}
where, the denominator of propagators are expressed as
\begin{eqnarray}
G_1&=& 2m_q^2+2t\cdot t^\prime,\nonumber\\
G_2&=& 2 s\cdot s^\prime,\nonumber\\
G_3&=& 2m_Q^2-2s\cdot p^\prime,\nonumber\\
G_4&=&2s^\prime\cdot t^\prime,\\
G_5&=& 2m_q^2+2t\cdot t^\prime,\nonumber\\
G_6&=& -2p^\prime\cdot s^\prime.\nonumber
\end{eqnarray}
Next, using the kinematics (\ref{kinematic}) we put the dot products of the relevant four-vectors in the following form
\begin{eqnarray}\label{mohsen}
2 p^\prime\cdot t^\prime&=&\frac{2}{1-z}(m_q^2+\frac{k_T^2}{4})+\frac{1-z}{2}(m_Q^2+k_T^2)-k_T^2,\nonumber\\
2 t\cdot t^\prime&=&\frac{2m_q z}{M(1-z)}(m_q^2+\frac{k_T^2}{4})+\frac{1-z}{2z}m_q M,\nonumber\\
2 p^\prime\cdot s^\prime&=&\frac{k_T^2}{2(1-z)}+\frac{1-z}{2}(m_Q^2+k_T^2)-k_T^2,\nonumber\\
2 s\cdot t^\prime&=&\frac{2zm_Q}{M(1-z)}(m_q^2+\frac{k_T^2}{4})+\frac{1-z}{2z}m_Q M,\nonumber\\
2 s\cdot p^\prime&=&\frac{m_Q M}{z}+\frac{zm_Q}{M}(m_Q^2+k_T^2),\nonumber\\
2 s\cdot s^\prime&=&\frac{z m_Q}{2M(1-z)}k_T^2+\frac{1-z}{2z} m_Q M,\\
2 t\cdot s^\prime&=&\frac{z m_q}{2M(1-z)}k_T^2+\frac{1-z}{2z}m_q M,\nonumber\\
2 t\cdot p^\prime&=&\frac{m_q M}{z}+\frac{m_q z}{M}(m_Q^2+k_T^2),\nonumber\\
2 s\cdot s &=& 2 p^\prime \cdot p^\prime= 2 m_Q^2,\nonumber\\
2 t\cdot t &=& 2 t^\prime \cdot t^\prime= 2 m_q^2,\nonumber\\
2 s\cdot t&=& 2m_q m_Q,\nonumber\\
2 s^\prime\cdot t^\prime&=&m_q^2,\nonumber\\
s^\prime\cdot s^\prime &=& 0.\nonumber
\end{eqnarray}
Substituting (\ref{base}) and (\ref{forth}) in (\ref{first}) and carrying out the necessary integrations, the fragmentation function $D_{Q\rightarrow M}$ reads
\begin{eqnarray}\label{hehe}
D_{Q\rightarrow M}^{Real}(z, \mu_0)=\frac{A^2\alpha_s^3}{3}C_F^2\int\frac{d^3 t^\prime d^3 s^\prime}{t_0^\prime s_0^\prime}
\int\frac{\sum_{i, j=1}^4\Gamma_i\cdot\Gamma_j^\star}{\bar{P}_0 p_0^\prime D_0^2}
\delta^3(\vec{\bar P}+
\vec{t^\prime}+\vec{s^\prime}-\vec{p^\prime})d^3\vec{\bar P},
\end{eqnarray}
where $A=\pi^{3/2}f_M m_q m_Q$ and the factor $D_0=\bar P_0+
t_0^\prime+s_0^\prime-p_0^\prime$ is the energy denominator.\\
To proceed one needs to determine the phase space integrations in (\ref{hehe}). Then we start with the following integral
\begin{eqnarray}\label{ayda1}
&&\int \frac{d^3\vec{\bar P}\delta^3(\vec{\bar P}+
\vec{t^\prime}+\vec{s^\prime}-\vec{p^\prime})}{p_0^\prime\bar{P}_0 D_0^2}=\frac{z}{G^2(z)},
\end{eqnarray}
where  $G(z)=M^2-m_Q^2-m_q^2-2t^\prime\cdot s^\prime+2p^\prime\cdot t^\prime+2p^\prime\cdot s^\prime$.
Considering the dot products of the four-vectors (\ref{mohsen}),  it is  simplified as 
\begin{eqnarray}
G(z)=M^2+\frac{z}{1-z}\big\{zk_T^2+2m_q^2-(1-z)m_Q^2\big\}.
\end{eqnarray}
For simplicity, we also assume that
the emitted gluon and outgoing light quark  $\it{q}$ move almost in the same direction.
This assumption is justified by the  fact that the very high momentum of the initial heavy quark is predominantly carried in the forward direction. Due to momentum conservation, the total transverse momentum of the emitted gluon and light quark will be identical to the transverse momentum of the initial heavy quark. Therefore, we have $t_T^\prime\approx s_T^\prime= k_T/2$.
According to the Suzuki's model, and for simplicity, we also replace the transverse momentum integrations by their average values as
\begin{eqnarray}\label{ayda2}
\int d^3 t^\prime\frac{f(z,t_T^{\prime 2})}{t_0^\prime}\approx  f(z, \frac{1}{4}\left\langle k_T^{2}\right\rangle),\quad\quad
\int d^3 s^\prime\frac{H(z,s_T^{\prime 2})}{s_0^\prime}\approx H(z, \frac{1}{4}\left\langle k_T^{2}\right\rangle),
\end{eqnarray}
where $ \left\langle k_T^2 \right\rangle$ is a free parameter which can be specified phenomenologically. Substituting all in (\ref{first}), we obtain the fragmentation function for the $Q\rightarrow M$ as follows
\begin{eqnarray}\label{lasttt}
D_{Q\rightarrow M}^{Real}(z, \mu_0)=\frac{A^2\alpha_s^3}{3}C_F^2\frac{z}{G^2(z)}\sum_{i,j=1}^4\Gamma_i\cdot\Gamma_j^\star.
\end{eqnarray}
The next step will be to compute $\Gamma_i\cdot\Gamma_j^\star$  and sum (or average) over gluon and quark polarization states.
At first, we calculate the contribution of  the forth Feynman diagram (Fig.~\ref{nlo}D)   to the radiative corrections by considering $\Gamma_4$ in (\ref{fifth}). Then one has
\begin{eqnarray}
\Gamma_4&=&\frac{-i}{G_7(q_4^{\prime 2}) G_8(q_4^2)}
\Big\{\bar{u}(s,r_2)\gamma^\kappa\big[(-2s^\prime\cdot \epsilon^\star-q^\prime\cdot \epsilon^\star)+(2q^\prime\cdot \epsilon^\star+s^\prime\cdot \epsilon^\star)+(s^\prime\cdot \epsilon^\star-q^\prime\cdot \epsilon^\star)\big]u(p^\prime,r_1)\Big\}\times\nonumber\\
&&\hspace{2.25cm}\Big\{\bar{u}(t^\prime,r_4)\gamma_\nu v(t,r_3)\Big\}=0,
\end{eqnarray}
and for the remaining contributions we have
\begin{eqnarray}
\sum_{r_i}\Gamma_1\cdot\Gamma_1^\star&=&-\frac{1}{G_1^2G_2^2}Tr\bigg\{(\displaystyle{\not}s+m_Q)
\gamma^\kappa(\displaystyle{\not}s+\displaystyle{\not}s^\prime+m_Q)\gamma^\mu(\displaystyle{\not}{p}^\prime+m_Q)
\gamma^\nu(\displaystyle{\not}s+\displaystyle{\not}s^\prime+m_Q)\gamma_\kappa\bigg\}\times Tr\bigg\{(\displaystyle{\not}{t}^\prime+m_q)\gamma_\mu(\displaystyle{\not} t-m_q)\gamma_\nu\bigg\},\nonumber\\
\sum_{r_i}\Gamma_2\cdot\Gamma_2^\star&=&-\frac{1}{G_3^2G_4^2}Tr\bigg\{(\displaystyle{\not}s+m_Q)
\gamma^\mu(\displaystyle{\not}{p}^\prime+m_Q)\gamma^\nu\bigg\}\times Tr\bigg\{(\displaystyle{\not}{t}^\prime+m_q)
\gamma^\kappa(\displaystyle{\not}{t}^\prime+
\displaystyle{\not}{s}^\prime+m_q)\gamma_\mu(\displaystyle{\not}t-m_q)\gamma_\nu (\displaystyle{\not}{t}^\prime+
\displaystyle{\not}{s}^\prime+m_q)\gamma_\kappa\bigg\},\nonumber\\
\sum_{r_i}\Gamma_3\cdot\Gamma_3^\star&=&-\frac{1}{G_5^2G_6^2}Tr\bigg\{(\displaystyle{\not}s+m_Q)
\gamma^\mu(\displaystyle{\not}p^\prime-\displaystyle{\not}s^\prime+m_Q)\gamma^\kappa(\displaystyle{\not}{p}^\prime+m_Q)
\gamma_\kappa(\displaystyle{\not}p^\prime-\displaystyle{\not}s^\prime+m_Q)\gamma^\nu\bigg\}\times Tr\bigg\{(\displaystyle{\not}{t}^\prime+m_q)\gamma_\mu(\displaystyle{\not} t-m_q)\gamma_\nu\bigg\},\nonumber\\
\sum_{r_i}\Gamma_1\cdot\Gamma_2^\star&=&-\frac{1}{G_1G_2G_3G_4}Tr\bigg\{(\displaystyle{\not}s+m_Q)
\gamma^\kappa(\displaystyle{\not}s+\displaystyle{\not}s^\prime+m_Q)\gamma^\mu(\displaystyle{\not}{p}^\prime+m_Q)
\gamma^\nu\bigg\}\nonumber\\
&&\hspace{2cm}\times Tr\bigg\{(\displaystyle{\not}t^\prime+m_q)\gamma_\mu(\displaystyle{\not}{t}-m_q)
\gamma_\nu(\displaystyle{\not}s^\prime+\displaystyle{\not}t^\prime+m_q)\gamma_\kappa\bigg\},\\
\sum_{r_i}\Gamma_1\cdot\Gamma_3^\star&=&-\frac{1}{G_1G_2G_5G_6}Tr\bigg\{(\displaystyle{\not}s+m_Q)
\gamma^\kappa(\displaystyle{\not}s+\displaystyle{\not}s^\prime+m_Q)\gamma^\mu(\displaystyle{\not}{p}^\prime+m_Q)
\gamma_\kappa(\displaystyle{\not}{p}^\prime-\displaystyle{\not}{s}^\prime+m_Q)\gamma^\nu\bigg\}\nonumber\\
&&\hspace{2cm}\times Tr\bigg\{(\displaystyle{\not}t^\prime+m_q)\gamma_\mu(\displaystyle{\not}{t}-m_q)\gamma_\nu\bigg\},\nonumber\\
\sum_{r_i}\Gamma_2\cdot\Gamma_3^\star&=&-\frac{1}{G_3G_4G_5G_6}Tr\bigg\{(\displaystyle{\not}s+m_Q)
\gamma^\mu(\displaystyle{\not}p^\prime+m_Q)\gamma_\kappa(\displaystyle{\not}{p}^\prime-\displaystyle{\not}{s}^\prime+m_Q)\gamma^\nu\bigg\}\nonumber\\
&&\hspace{2cm}\times Tr\bigg\{(\displaystyle{\not}t^\prime+m_q)\gamma^\kappa(\displaystyle{\not}t^\prime+
\displaystyle{\not}s^\prime+m_q)\gamma_\mu(\displaystyle{\not}t-m_q)\gamma_\nu\bigg\}.\nonumber
\end{eqnarray}
In calculating the above terms we used the identities  $\sum_r u(p,r) \bar{u}(p,r)=\displaystyle{\not}p+m$ and 
$\sum_r v(p,r) \bar{v}(p,r)=\displaystyle{\not}p-m$ for the polarization sums. There is a similar trick
for summing over gluon polarization vectors. The correct prescription is to make
the replacement: $\sum_{\lambda}\epsilon_\mu(\lambda)\epsilon_\nu^\star(\lambda)\rightarrow -g_{\mu\nu}$ (see Sec.5.5 of Ref.~\cite{Peskin:1995ev}).  

Considering the Dirac algebra and using the dot products of four-momenta, these expressions can be simplified as
\begin{eqnarray}\label{ali}
\sum_{r_i}\Gamma_1\cdot\Gamma_1^\star&=&\frac{2m_qm_Q}{G_1^2G_2^2}\bigg\{\frac{m_Q^4}{z^2}(z-1)(19z^2-6z+3)
-\frac{m_Q^2 k_T^2}{1-z}(21z^2-10z+5)-\frac{z^2k_T^4}{1-z}+\frac{z^4 k_T^6}{m_Q^2(1-z)^3}\bigg\},\nonumber\\
\sum_{r_i}\Gamma_2\cdot\Gamma_2^\star&=&\frac{24m_q^3}{G_3^2G_4^2}\bigg\{\frac{m_Q^3}{z^2}(z-1)(3z^2-2z+1)
-2\frac{m_Q k_T^2}{1-z}(2z^2-2z+1)-\frac{z^2k_T^4}{m_Q(1-z)}\bigg\},\nonumber\\
\sum_{r_i}\Gamma_3\cdot\Gamma_3^\star&=&\frac{2m_q}{G_5^2G_6^2}\bigg\{\frac{m_Q^5}{z^2}(z-1)(19z^2-6z+3)
-\frac{m_Q^3 k_T^2}{1-z}(21z^2-10z+5)-\frac{z^2k_T^4}{1-z}m_Q+\frac{z^4 k_T^6}{m_Q(1-z)^3}\bigg\},\nonumber\\
\sum_{r_i}\Gamma_1\cdot\Gamma_2^\star&=&\frac{-8}{G_1 G_2 G_3 G_4}\frac{m_q}{z^3(1-z)^2m_Q}\bigg[
(1-z)^2m_Q^2+z^2k_T^2\bigg]^2\bigg[z^2k_T^2+m_Q^2(1-2z+3z^2)\bigg],\nonumber\\
\sum_{r_i}\Gamma_1\cdot\Gamma_3^\star&=&\frac{4m_q}{G_1G_2G_5G_6}\bigg\{\frac{m_Q^5(z-1)}{z^3}
(5z^4-2z^3+6z^2-2z+1)-\frac{m_Q^3k_T^2}{z(1-z)}(12z^4-19z^3+21z^2-9z+3)-\nonumber\\
&&\frac{zm_Qk_T^4}{1-z}(9z^2-6z+3)+\frac{z^3(2z-1)k_T^6}{m_Q(1-z)^2}\bigg\},\nonumber\\
\sum_{r_i}\Gamma_2\cdot\Gamma_3^\star&=&\frac{-8}{G_3G_4G_5G_6}\frac{m_q}{z^2(1-z)^2m_Q}\bigg[
z^2k_T^2+m_Q^2(1-2z+3z^2)\bigg]\bigg[z^2k_T^2+(1-z)^2m_Q^2\bigg]^2.
\end{eqnarray}
\begin{figure}
	\begin{center}
		\includegraphics[width=0.5\linewidth,bb=88 530 452 700]{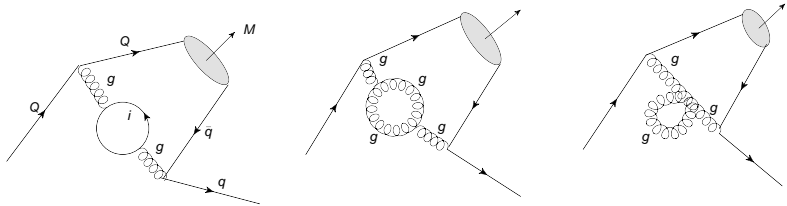}
		\caption{\label{virtual}%
			Virtual gluon contributions to the $Q\rightarrow M(Q\bar{q})+q$ at NLO.}
	\end{center}
\end{figure}
At NLO approximation, in addition to the real gluon radiative corrections there are some Feynman diagrams related to the virtual  corrections.
This class of contributing diagrams interferes with the LO amplitude.
In fact, the NLO full  amplitude is the sum of the amplitudes of the Born term ($\Gamma^{LO}$),
virtual one-loop ($\Gamma^{Loop}$), and the real contributions ($\Gamma^{Real}$),
\begin{eqnarray}
	\Gamma^{NLO}=\Gamma^{LO}+\Gamma^{Loop}+\Gamma^{Real}.
\end{eqnarray}
The QCD NLO contributions result from the square of the amplitudes:
 $|\Gamma^{Born}|^2=\Gamma^{LO}\cdot\Gamma^{LO\star}$ so its related  fragmentation function  is of order $\alpha_s^2$ (\ref{last}), $|\Gamma^{Vir}|^2=2Re(\Gamma^{LO}\cdot\Gamma^{Loop\star})$ and $|\Gamma^{Real}|^2=\Gamma^{Real}\cdot\Gamma^{Real\star}$ (\ref{ali}) so that the NLO fragmentation function is of order $\alpha_s^3$ (\ref{lasttt}).
The Feynman diagrams related to the virtual  gluon radiative corrections are classified into two classes. The first class of contributing diagrams includes  the fermion loop diagram, the three-gluon vertex and a four-gluon vertex. These are shown in Fig.~\ref{virtual}. 
It is shown that these virtual contributions interlock in an essential way.
In general, Feynman diagrams with $n$ loops typically contain correction terms proportional to $(\alpha_s\log(Q^2/\Lambda^2))^n$, where $\Lambda$ is a renormalization scale. Fortunately, we can absorb these corrections into the lowest-order terms by using the renormalization
group equations. In other words, their effect  is  to modify the gluon propagator by replacing the fixed renormalized coupling with a running coupling constant.
By solving the renormalization group equations,
the one-loop corrections shown lead to evolve the QCD coupling constant at the energy Q, as \cite{Catani:1998tm}
\begin{eqnarray}\label{akh}
	\alpha_s(Q^2)=\frac{2\pi}{b_0\log\frac{Q}{\Lambda}},
\end{eqnarray}
with $b_0=11-2n_f/3$, where $n_f$ refers to the active quark flavor numbers. In the above equation, $\Lambda$ is a typical QCD scale which shows the border
between the perturbative and nonperturbative regimes of QCD. 
In practice the value of $\Lambda$ is ambiguous and  is  determined by experiments. Experimental measurements of the rate of $e^-e^+$ reaction and others yield a value of $\Lambda\approx 231$ MeV \cite{Nakamura:2010zzi} so the QCD perturbation theory is valid only when  $Q$ is somewhat larger than this, say above $Q=1$ GeV, where $\alpha_s(Q)\approx 0.4$.\\
Besides these 1PI diagrams, there are also three $\it{tadpole}$ diagrams;
one-loop diagrams with a propagator that connects back to its originating vertex.
It is shown that these automatically vanish.

The second class of virtual corrections includes the gluon-quark loops on the incoming or the outgoing quark legs. Generally, these amplitudes need to be considered and include in order to maintain the infrared stability of the overall result. Indeed, these virtual corrections consist of both
infrared (IR) and ultraviolet (UV) singularities where the
UV-divergences appear when the integration region of the internal
momentum of the virtual gluon goes to infinity and
the IR-divergences arise from the soft-gluon singularities.
All UV-singularities are canceled after
summing all virtual contributions up, whereas the IR
singularities are remaining. The real gluon radiative corrections also include IR-divergences which arise
from the soft- and collinear gluon emissions.
According to the Lee-Nauenberg theorem, after summing all radiative corrections up the IR-singularities cancel each other and the final result
is free of all singularities. More  details can be found 
in our previous works \cite{MoosaviNejad:2012ju,Nejad:2016epx,Nejad:2015pca,Nejad:2014sla,Nejad:2013fba,Kniehl:2012mn,MoosaviNejad:2011yp}, where we calculated the decay rate of top quarks at NLO by working at dimensional regularization scheme. 

Note that in the Suzuki's model, to compute the contribution of the real corrections into the fragmentation function we do not integrate over the momentum of the emitted real gluon and instead, we replace the  gluon  momentum integration by its average value, see Eq.~(\ref{ayda2}). Therefore, by this simplification on one side we shall not deal with the IR-singularities in the real gluon radiative corrections (\ref{ali}) and on the other side the contribution of the  virtual corrections can be ignored. We checked that the contribution of the virtual gluon corrections into the QCD amplitude $T_H$ (\ref{forth}) is small and then their corresponding FFs are  tiny. Specifically, this point is confirmed in Fig.~\ref{belle}.

\section{Numerical analysis}
\label{three}
\begin{figure}
	\begin{center}
		\includegraphics[width=0.5\linewidth]{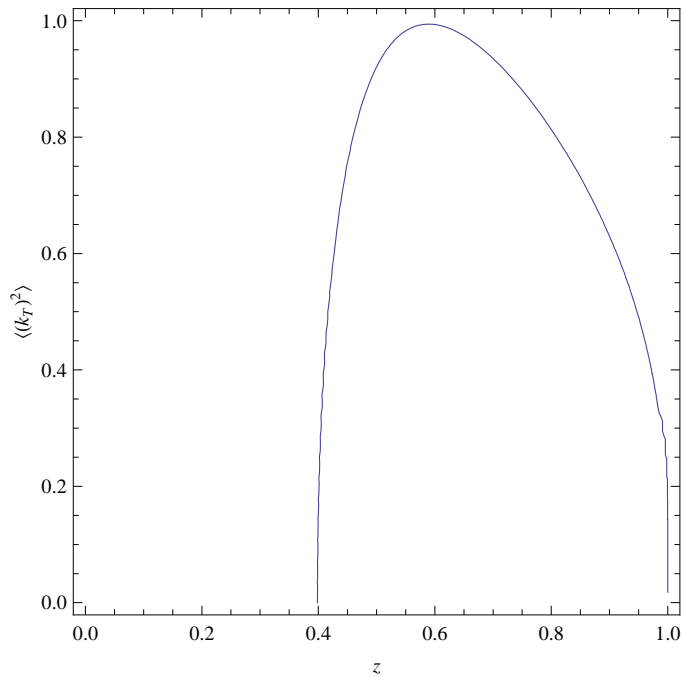}
		\caption{\label{fig}%
			The behavior of $\left\langle k_T^2 \right\rangle$ as a function of $z$ when $D(z)$ is normalized to unity.
			}
	\end{center}
\end{figure}
\begin{figure}
	\begin{center}
		\includegraphics[width=0.6\linewidth]{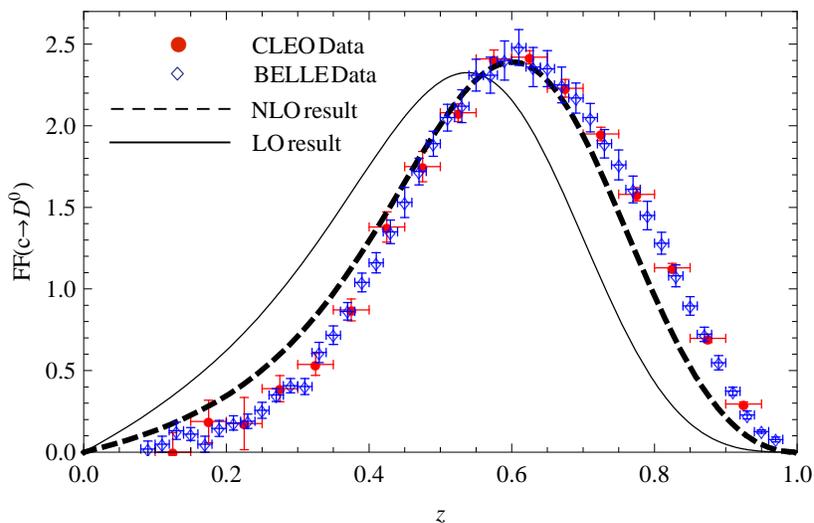}
		\caption{\label{belle}%
			The pQCD FF of $c\rightarrow D^0$ at LO (solid line) and NLO (dashed line)  approximations. The fragmentation scale is $\mu_0=m_c$ and  we set $\left\langle k_T^2 \right\rangle=1$ GeV. The theoretical results are also compared with data from BELLE \cite{belle} and CLEO \cite{cleo}.}
	\end{center}
\end{figure}
\begin{figure}
	\begin{center}
		\includegraphics[width=0.6\linewidth]{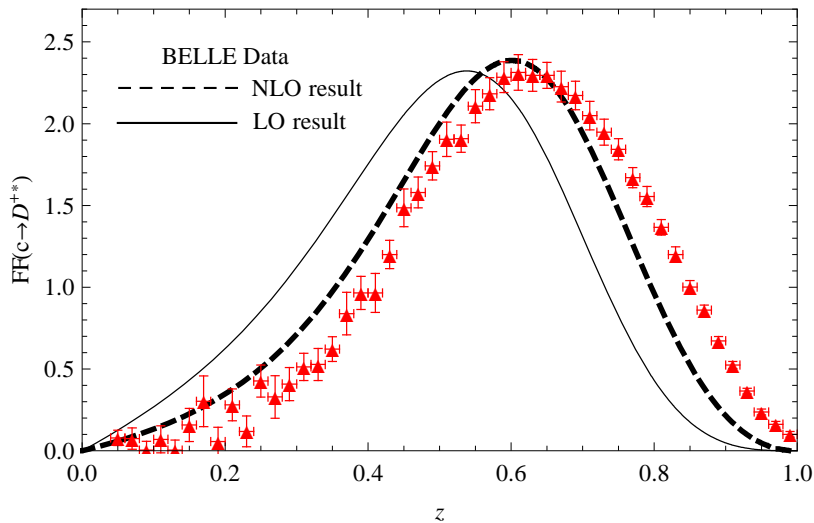}
		\caption{\label{dplus}%
			Comparison of pQCD FF with data from BELLE \cite{belle} on $D^{+\star}$ production at the initial scale $\mu_0=m_c$, considering  LO (solid line) and NLO (dashed line)  approximations.}
	\end{center}
\end{figure}
We are now in a position to present our numerical results for the heavy quark fragmentation function.  Our result at the leading-order approximation is shown in (\ref{last}) and the result at NLO is obtained as follows
\begin{eqnarray}\label{final}
D_{Q\rightarrow M}^{NLO}(z,\mu_0)=N(D^{LO}_{Q\rightarrow M}+D^{Real}_{Q\rightarrow M}),
\end{eqnarray}
where $N$ is obtained through the normalization condition
$\int_0^1 D_Q^M(z, \mu_0) dz=1$  \cite{Amiri:1986zv,Suzuki:1985up}, and
the $D^{LO}$ and $D^{Real}$ are presented  in (\ref{last}) and (\ref{lasttt}), respectively. 
Generally, the fragmentation function $D_{Q\rightarrow M}(z,\mu_0)$ depends on both the 
fragmentation parameter $z=E_M/E_Q (0\leq z\leq 1)$ and the fragmentation  scale $\mu$.
This scale is normally arbitrary, but in  a high energy process of electron-positron annihilation
where a jet is produced with transverse momentum $ k_T$,
large logarithms of $k_T/\mu$ in the partonic cross section of the process $e^+e^-\rightarrow Q\bar Q\rightarrow M(Q\bar q)+X$
can be avoided by choosing  $\mu$  on the order of $ k_T$.
Also, the $z$ dependence of the FF is not yet calculable at each desired scale. However, once
they are computed at some initial fragmentation scale $\mu_0$, their $\mu$ evolution is specified by
the  DGLAP evolution equations \cite{dglap}.
Therefore, the function (\ref{final}) should be regarded as a model for the heavy quark FF at the scale $\mu_0$ of 
order $m_Q$ and the $D_{Q\rightarrow M}(z,\mu)$
at larger scales can be obtained  by solving DGLAP equations.

Here, as an example, 
we consider the fragmentation  of $\it{c}$-quark into $D^0$- and $D^{+\star}$-mesons 
with the constituent quark structures $|D^0>=|c\bar u>$ and $|D^{+\star}>=|c\bar d>$ using  $m_Q=m_c=1.3$ GeV, $m_u=3$ MeV,  $m_d=5$ MeV and  
$f_M=0.22$ GeV  \cite{Nakamura:2010zzi}. In this work we also consider
 $\alpha_s(m_c)=0.38\pm 0.03$ adjusted  such that $\alpha_s(m_Z)=0.1184$ with $m_Z=91.18$ GeV.\\
 It is worth mentioning here that the average transverse momentum is not a constant but a function of the fragmentation parameter $z$.  Our prediction for $z$ dependence of this quantity is shown in Fig.~\ref{fig}. This plot justifies the fact that the choice of $\left\langle k_T^2 \right\rangle=1$ GeV is an extreme value for this quantity and any lower value will produce the peak even at higher-z regions.
In Figs.~\ref{belle} and \ref{dplus}, the behavior of $D^0$ and $D^{+\star}$ FFs at the initial scale $\mu_0=m_c$ is shown for the LO (solid line) and NLO (dashed line) approximations. For comparison, data from BELLE \cite{belle} and CLEO \cite{cleo} are also shown, see also Fig.~3 from Ref.~\cite{Corcella:2007tg}. As is seen, there is 
reliable consistency between our analytic result at NLO and experimental data. However, 
we may also  think of other effects such as the Fermi motion of constituent quarks,
the meson relativistic wave function \cite{MoosaviNejad:2016scq} and the effects of meson mass and so on, which can make  a better agreement with experimental data. Concerning the data shown in Figs.~\ref{belle} and \ref{dplus} we shall discuss, in this section, later.

\begin{figure}
	\begin{center}
		\includegraphics[width=0.6\linewidth]{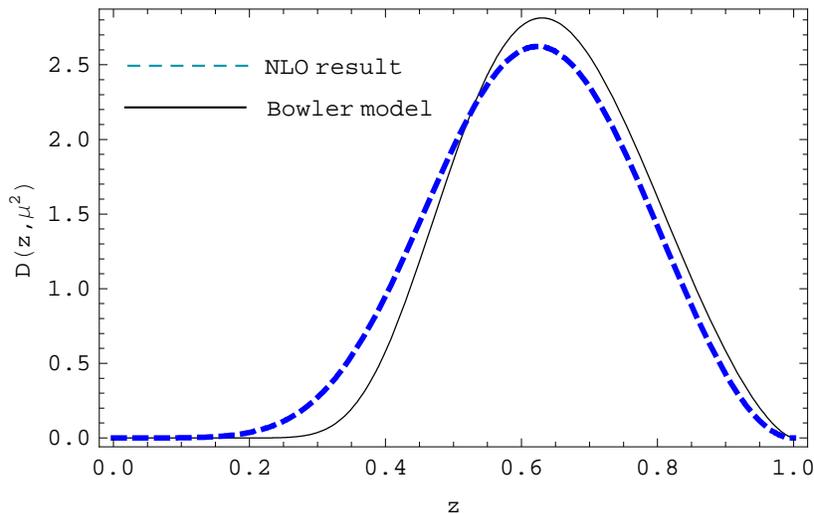}
		\caption{\label{bowler}%
			The $D_c^{D^0}$ FF at the initial scale $\mu_0=m_c$ as a function of $z$ in the pQCD approach (dashed line)
			and Bowler model (solid line). The result in the pQCD approach is obtained at NLO approximation.}
	\end{center}
\end{figure}

Besides the theoretical schemes  
there is another current approach to determine the FFs; {\it phenomenological} approach. This is 
based on  data analyzing where the FFs are mainly determined by hadron production data of electron-positron annihilation,  hadron-hadron and lepton-hadron scattering processes. However, among  all the $e^+e^-$ annihilation provides a clean environment to determine the fragmentation densities, specifically one does not need to consider the parton distribution functions (PDFs) of initial hadrons. In this scheme the Collin's factorization theorem of the QCD improved parton model \cite{Collins:1998rz} is an important tool to study this process. According to this theorem,
in the high energy scattering $e^+e^-\rightarrow H+X$ the cross section of hadron production  is described by the convolution of partonic hard scattering cross sections $d\sigma_i(e^+e^-\rightarrow i\bar{i})/dx_i$, which are calculable in perturbative QCD \cite{Webber:1983if}, and the realistic and  nonperturbative FFs $D_i^H$ describing the transition of a parton into an outgoing hadron ($i/\bar{i}\rightarrow H$), i.e.
\begin{eqnarray}\label{monaa}
\frac{1}{\sigma_{tot}}\frac{d}{dz}\sigma(e^+e^-\rightarrow HX)=\sum_i \int_z^1\frac{dx_i}{x_i} D_i^H(\frac{z}{x_i}, \mu)\frac{1}{\sigma_{tot}}\frac{d\sigma_i}{dx_i}(x_i, \mu),
\end{eqnarray}
where, the momentum fraction $z$ is defined as
$z=E_H/E_Q=2 E_H/\sqrt{s}$ where $E_H$ is the energy of hadron and $s$ is the square of total $e^+e^-$ center-of-mass energy ($E_Q=\sqrt{s}/2$).
In (\ref{monaa}), $x_i$ is defined as $x_i=2E_i/\sqrt{s}$ and 
$X$ stands for the unobserved jets and  $\sigma_{tot}$  is the total partonic  cross section at NLO \cite{Kneesch:2007ey}.
In this scheme, the FFs are parameterized in terms of a number of free parameters which are determined by an $\chi^2$ analysis of the
$e^+e^-$ annihilation data at the scale $\mu^2=Q^2$ where $Q^2$ is the squared center-of-mass energy.
These parameterizations should include some restrictions. For example, they must be zero at $z=0$ and $z=1$.
Various phenomenological models like Peterson model \cite{Peterson:1982ak}, Lund model \cite{Andersson:1983ia},
Cascade model \cite{Webber:1983if} and etc., have been developed to describe the FFs.\\
In Ref.~\cite{Kneesch:2007ey},  authors
computed the FFs of $D^0, D^+$ and $D^{\star +}$ mesons
through a global fit to electron-positron data from
the BELLE, CLEO, ALEPH, and OPAL collaborations.
According to the Bowler model  \cite{Bowler:1981sb}, authors have parameterized the $z$ distributions of the charm quark FF at its starting scale $\mu_0=m_c$, as
\begin{eqnarray}\label{bw}
D_q^{H_c}(z,\mu_0)=Az^{-(1+\gamma^2)}(1-z)^a e^{-\gamma^2/z},
\end{eqnarray}
with three free parameters.  Their result for $D^0$-meson reads $A=3.43\times 10^4$, $a=1.48$ and $\gamma=2.80$ with the value of $\chi^2=0.789$ achieved. In Fig.~\ref{bowler}, using  (\ref{final}) the behavior of $D^0$ FF at the starting scale $\mu_0=m_c$ is  compared  with  the Bowler model,  as a well-known  phenomenological model. 
Since to obtain the constant $N$  (\ref{final}) we have used the normalization condition
then to compare our result with the Bowler model, the fragmentation function in the Bowler model should be divided by the $c\rightarrow D^0$
branching fraction $B(m_c)=0.634$ \cite{Kneesch:2007ey}. The branching fraction  is defined as $B_c(\mu)=\int_{z_{cut}}^1dz D(z,\mu^2)$
where the cut $z_{cut}$ excludes the $z$ range in which  the result is not valid.
As Fig.~\ref{bowler} shows our result at NLO is in reliable consistency  with the phenomenological model. In this comparison we set $m_c=1.5$ GeV as in \cite{Kneesch:2007ey}.

Concerning the data shown in Figs.~\ref{belle} and \ref{dplus}, it should be noted that according to the definition of  FF presented in \cite{Amiri:1986zv}, the FF $D_Q^M(z)$ is related to the differential  cross section for the inclusive meson production  as 
\begin{eqnarray}\label{aydaa2}
D_Q^H(z)=\frac{1}{\sigma}\frac{d}{dz}\sigma(e^-e^+\rightarrow Q\bar{Q}\rightarrow H+X)
\end{eqnarray}
with normalization condition $\int_0^1 D_Q^H(z) dz=1$.\\
In \cite{Amiri:1986zv}, authors have compared their results for the fragmentation functions of $c$- and $b$-quarks into $D$-
and $B$-mesons with various experimental data. In \cite{Suzuki:1977km}, Suzuki have compared the FF of $\pi$ with the inclusive antineutrino data in the process $\bar{\nu}+P\rightarrow \mu^++\pi^-+X$.
However, the mentioned authors (Suzuki \cite{Suzuki:1977km}, Amiri, Ji \cite{Amiri:1986zv} and {\it etc}) define the FFs as the full differential
hadron-level cross section, which is what is measured (\ref{aydaa2}), but
in phenomenological schemes (\ref{monaa}), one usually writes a hadronic cross section as a convolution of coefficient functions and fragmentation functions.
In fact, it is just a matter by definition and notation.
As long as one is consistent, both definitions are possible: according
to the definition, one can say that the experimental data are a differential cross section or a fragmentation function.
Similar issues hold for structure functions and parton distribution
functions.
In practice, it is also possible to show that both definitions are consistent at LO and at higher-orders they are equal approximately. According to the second definition, the cross section for $e^+e^-$ annihilation can be expressed as in (\ref{monaa}).\\
At LO, the Wilson coefficient functions are expressed as \cite{Kneesch:2007ey}
\begin{eqnarray}
\frac{1}{\sigma_{tot}}\frac{d\sigma_i}{dx_i}(x_i, \mu)=\delta(1-x_i)
\end{eqnarray}
then one has
\begin{eqnarray}
D_i^H(z, \mu)= \frac{1}{\sigma_{tot}}\frac{d}{dz}\sigma(e^+e^-\rightarrow HX),
\end{eqnarray}
which is the definition introduced by Amiri, Ji, Suzuki and {\it etc} (\ref{aydaa2}).
At NLO approximation, the Wilson coefficients read \cite{Kneesch:2007ey}
\begin{eqnarray}
\frac{1}{\sigma_{tot}}\frac{d\sigma_i}{dx_i}(x_i, \mu)=\delta(1-x_i)+\frac{\alpha_s}{2\pi}f(x_i, \mu),
\end{eqnarray}
so that in high energy $e^+e^-$ annihilation (the condition applied by Amiri and {\it etc}), the QCD coupling constant is tiny then  the definition is  approximately valid.

Besides the $c\rightarrow D^0/D^{\star +}$ FFs themselves, also their first moment is of phenomenological interest and subject
to experimental determination. It  corresponds to the average  fraction of energy that the $D^0/D^{\star +}$-mesons receive from
the $c$ quark, 
\begin{eqnarray}\label{ave}
\left\langle z \right\rangle_c (\mu)=\frac{1}{B_c(\mu)}\int_{z_{cut}}^1 dz z D_c(z,\mu^2),
\end{eqnarray}
where the cut  $z_{cut}=0.1$ excludes the problematic $z$
range where the formalism is not valid. On the other hand, as may be seen from Fig.~\ref{belle} there are no experimental data at $z<0.1$.
Our results for  the average energy fraction  is $\left\langle z \right\rangle (m_c)=0.48$ at LO and $\left\langle z \right\rangle (m_c)=0.45$ at NLO approximations. These results can be compared with the values quoted by BELLE, CLEO, ALEPH and OPAL which are listed  in \cite{Kneesch:2007ey}. Also, if one takes the Bowler model (\ref{bw}) with the values of free parameters presented above, the result would be $\left\langle z \right\rangle (m_c)=0.43$. There is  good consistency between our result
and the phenomenological  results, however one must keep in mind that experimental results naturally include
all orders and also contributions from gluon and light-quark fragmentation, while ours are evaluated at NLO.

\section{Conclusion}
\label{sec:four}

The dominant mechanism to produce hadronic bound states with large transverse momentum is fragmentation, that is the splitting of a high-energy parton into a hadronic state and other partons.
It is tempting to use the heavy-quark limits of the perturbative QCD fragmentation functions as
phenomenological models for the fragmentation of a heavy quark Q into heavy-light mesons $Q\bar{q}$, where $Q=c$ or $b$ and $q=u, d$, or $s$.
In this work, using the Suzuki's model we studied the perturbative QCD fragmentation functions for a heavy quark to fragment into  S-wave heavy-light mesons in the heavy-quark limit at NLO.
In this model, the nonperturbative aspect of the hadroproduction processes is emerged in the bound state of the meson
which is described by the wave function. 
Our result describes not only the $\it{z}$ dependence of the fragmentation probabilities, but also their dependence on the transverse momentum of the initial parton.
As a numerical example, we studied the initial FFs of c-quark to split into
S-wave $D^0/D^{+\star}$-mesons to leading order in $\alpha_s$ and next-to-leading order.
Specifically, we compared the LO and NLO FFs for $D^0/D^{+\star}$-mesons with available $e^-e^+$ annihilation data from BELLE \cite{belle} and CLEO \cite{cleo} and we found good agreement between the NLO result and experimental data. Our results are also compared with a well-known phenomenological model (Bowler model  \cite{Bowler:1981sb}) for the heavy-quark fragmentation and found  reliable consistency. The full agreement between our result and the experimental data can reach by considering some additional effects ignored in this work such as, the Fermi motion of constituent quarks, the meson wave function effects, etc.
It should be noted that, the next-to-next-to-leading order (NNLO) effects for the $c\rightarrow D$ FF will be of order $\alpha_s^4$ and these effects would not be expected to be sizable numerically. To include the NNLO effects, apart form the virtual corrections, one  needs to consider many real gluon Feynman diagrams  including the gluon, light and heavy quark propagators. Normally, we expect that the NNLO corrections would  increase the FF at $z$-large and decrease then at $z$-low to make a better fit with the data, as we had for the NLO effect, see Figs.~\ref{belle} and \ref{dplus}.

\end{document}